\documentclass[a4paper]{article}
\usepackage{listings}
\usepackage{xcolor}
\usepackage{multirow}
\usepackage{jheppub}  
\usepackage{dcolumn}
\usepackage[english]{babel}
\usepackage[utf8x]{inputenc}
\usepackage[T1]{fontenc}
\usepackage{palatino}
\usepackage{amsmath}
\usepackage{afterpage}
\usepackage{lineno}
\usepackage{graphicx, subcaption}
\usepackage[colorinlistoftodos]{todonotes}
\usepackage[colorlinks=true]{hyperref}
\usepackage{makeidx}
\newcommand{\be}{\begin{equation}}  
\newcommand{\ee}{\end{equation}}  
\newcommand{\bea}{\begin{eqnarray}}  
\newcommand{\eea}{\end{eqnarray}}  
\makeindex
\begin{document}
%\linenumbers
\vspace*{1.2cm}

\thispagestyle{empty}
\begin{center}
{\LARGE \bf The LHCspin project}

\par\vspace*{7mm}\par

{
\bigskip
\large \bf 
M. Santimaria\textsuperscript{1$\star$},
V. Carassiti\textsuperscript{2},
G. Ciullo\textsuperscript{2,3},
P. Di Nezza\textsuperscript{1},
P. Lenisa\textsuperscript{2,3},
S. Mariani\textsuperscript{4,5},
L. L. Pappalardo\textsuperscript{2,3} and
E. Steffens\textsuperscript{6}
}

\bigskip

\begin{center}
{\bf 1} INFN Laboratori Nazionali di Frascati, Frascati, Italy
\\
{\bf 2} INFN Ferrara, Italy
\\
{\bf 3} Department of Physics, University of Ferrara, Italy
\\
{\bf 4} INFN Firenze, Italy
\\
{\bf 5} Department of Physics, University of Firenze, Italy
\\
{\bf 6} Physics Dept., FAU Erlangen-Nurnberg, Erlangen, Germany
\\
* marco.santimaria@lnf.infn.it
\end{center}

\bigskip

{\it Presented at the Low-$x$ Workshop, Elba Island, Italy, September 27--October 1 2021}

\end{center}
\vspace*{1mm}

\section*{\center \small Abstract}
The goal of LHCspin is to develop, in the next few years, innovative solutions and cutting-edge technologies to access spin physics in polarised fixed-target collisions at high energy, exploring the unique kinematic regime offered by LHC and exploiting new final states by means of the LHCb detector.
The forward geometry of the LHCb spectrometer is perfectly suited for the reconstruction of particles produced in fixed-target collisions. 
This configuration, with centre of mass energies
ranging from $\sqrt{s_{\rm{NN}}}=115~\rm{GeV}$ in $p-p$ interactions to
$\sqrt{s_{\rm{NN}}}=72~\rm{GeV}$ in heavy ion collisions,
allows to cover a wide backward rapidity region, 
including the poorly explored high$-x$ regime.
With the instrumentation of the proposed target system, LHCb will become the first experiment simultaneously collecting unpolarised beam-beam collisions at $\sqrt{s_{\rm{NN}}}=14~\rm{TeV}$ and both unpolarised and polarised beam-target collisions.
The status of the project is presented along with a selection of physics opportunities.

\section{Introduction}
\label{sec:intro}
The LHC delivers proton and lead beams with an energy of $7~\rm{TeV}$ and $2.76~\rm{TeV}$ per nucleon, respectively, with world's highest intensity. A short run with xenon ions was also performed in 2017, while an oxygen beam is currently foreseen for the Run 3~\cite{Citron:2018lsq}.
Fixed-target proton-gas collisions occur at a centre of mass energy per nucleon of up to $115~\rm{GeV}$.
This corresponds to a centre of mass rapidity shift of
$y-y_{\rm{CM}} \approx \rm{arcsinh}(\sqrt{E_{\rm{N}} / 2M_{\rm{N}}})=4.8$, so that the LHCb acceptance $(2<\eta<5)$ covers the backward and central rapidities in the centre of mass frame.
Such a coverage offers an unprecedented opportunity to investigate partons carrying a large fraction of the target nucleon momentum, i.e.~large Bjorken$-x$ values, corresponding to large and negative Feynman$-x$ values\footnote{$x_{\rm{F}} \approx x_1 - x_2$ where $x_1$ and $x_2$ are the Bjorken$-x$ values of the beam and target nucleon, respectively.}.

\begin{figure}[ht]
\centering
\includegraphics[width=0.9\textwidth]{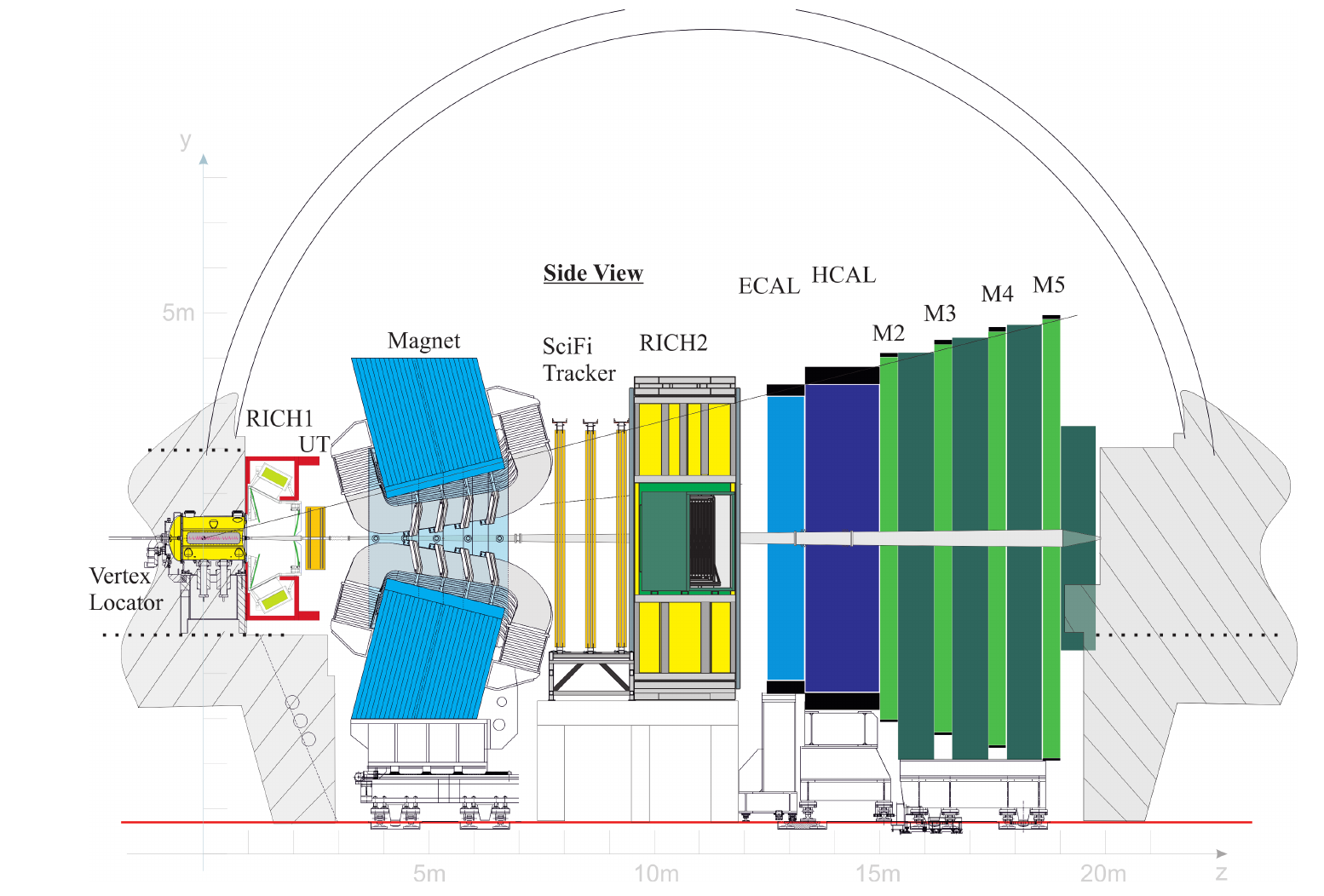}
\caption{The Run 3 LHCb detector.}
\label{fig:lhcb}
\end{figure}

The LHCb detector~\cite{Alves:2008zz} is a general-purpose forward spectrometer specialised in detecting hadrons containing $c$ and $b$ quarks, and the only LHC detector able to collect data in both collider and fixed-target mode.
It is fully instrumented in the $2<\eta<5$ region with a vertex locator (VELO), a tracking system, two Cherenkov detectors, electromagnetic and hadronic calorimeters and a muon detector.
Fig.~\ref{fig:lhcb} shows a scheme of the upgraded LHCb detector which is currently being installed for the Run 3, starting in 2022.

\begin{figure}[ht]
\centering
\includegraphics[width=0.99\textwidth]{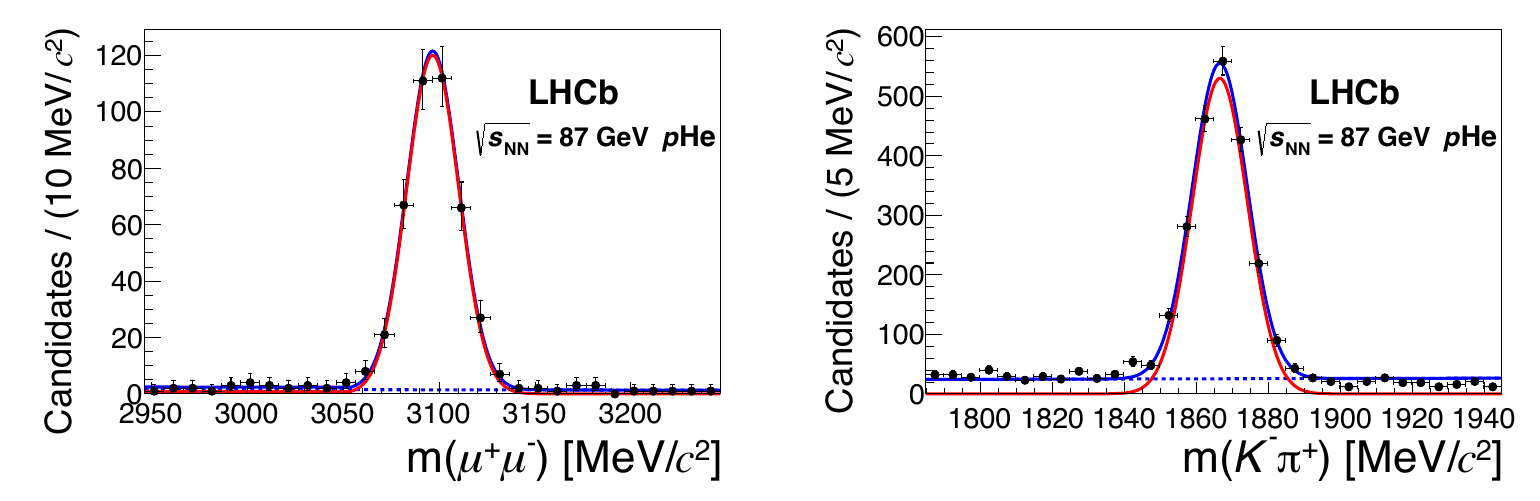}
\caption{$J/\psi\to\mu^+\mu^-$ (left) and $D^0\to K^-\pi^+$ (right) SMOG samples from~\cite{LHCb:2018ygc}.}
\label{fig:smog_phe}
\end{figure}

The fixed-target physics program at LHCb is active since the installation of the SMOG (System for Measuring the Overlap with Gas) device~\cite{Aaij:2014ida}, enabling the injection of noble gases in the beam pipe section crossing the VELO detector at a pressure of $\mathcal{O}(10^{-7})~\rm{mbar}$.
Precise measurements of charm~\cite{LHCb:2018ygc} and antiproton~\cite{LHCb:2018jry} production have been published based on $p-\rm{Ar}$ and $p-\rm{He}$ fixed-target collisions. Fig.~\ref{fig:smog_phe} shows the high-quality and low-background samples collected in just one week of dedicated SMOG runs during Run 2.

\begin{figure}[ht]
\centering
\includegraphics[width=0.49\textwidth]{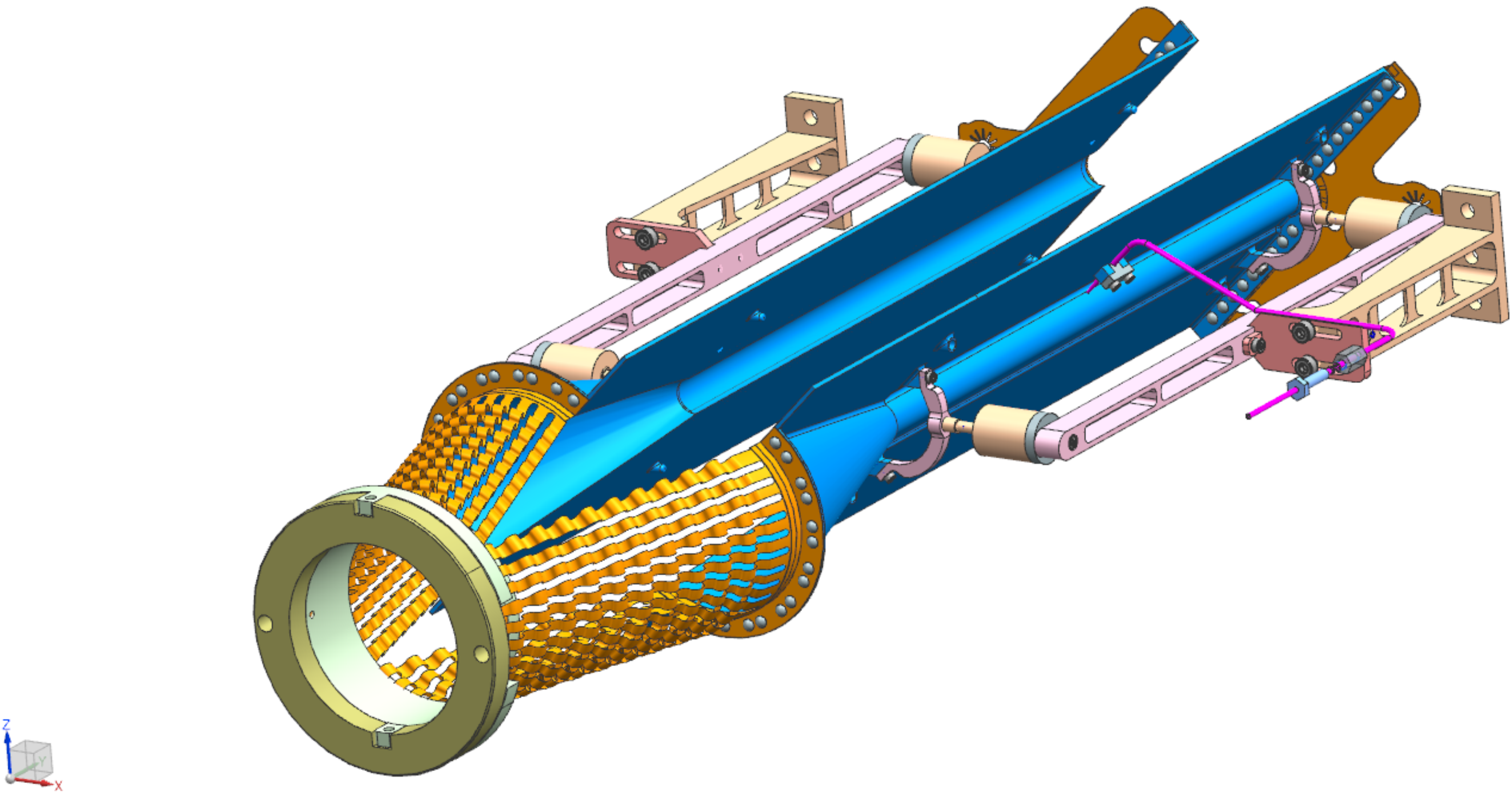}
\includegraphics[width=0.49\textwidth]{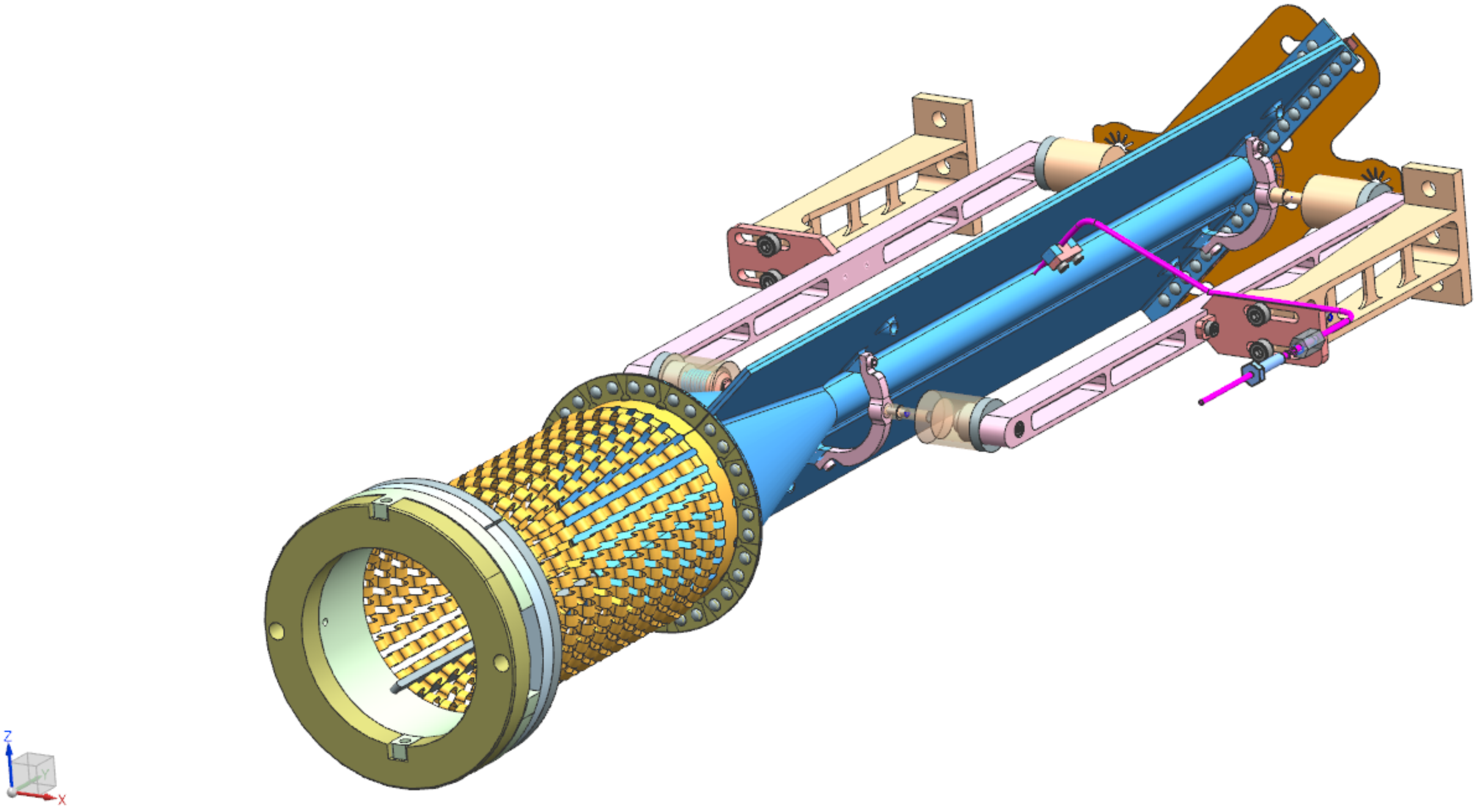}
\caption{The SMOG2 storage cell in the open (left) and closed (right) configuration.}
\label{fig:smog2}
\end{figure}

\begin{figure}[h]
\centering
\includegraphics[width=0.52\textwidth]{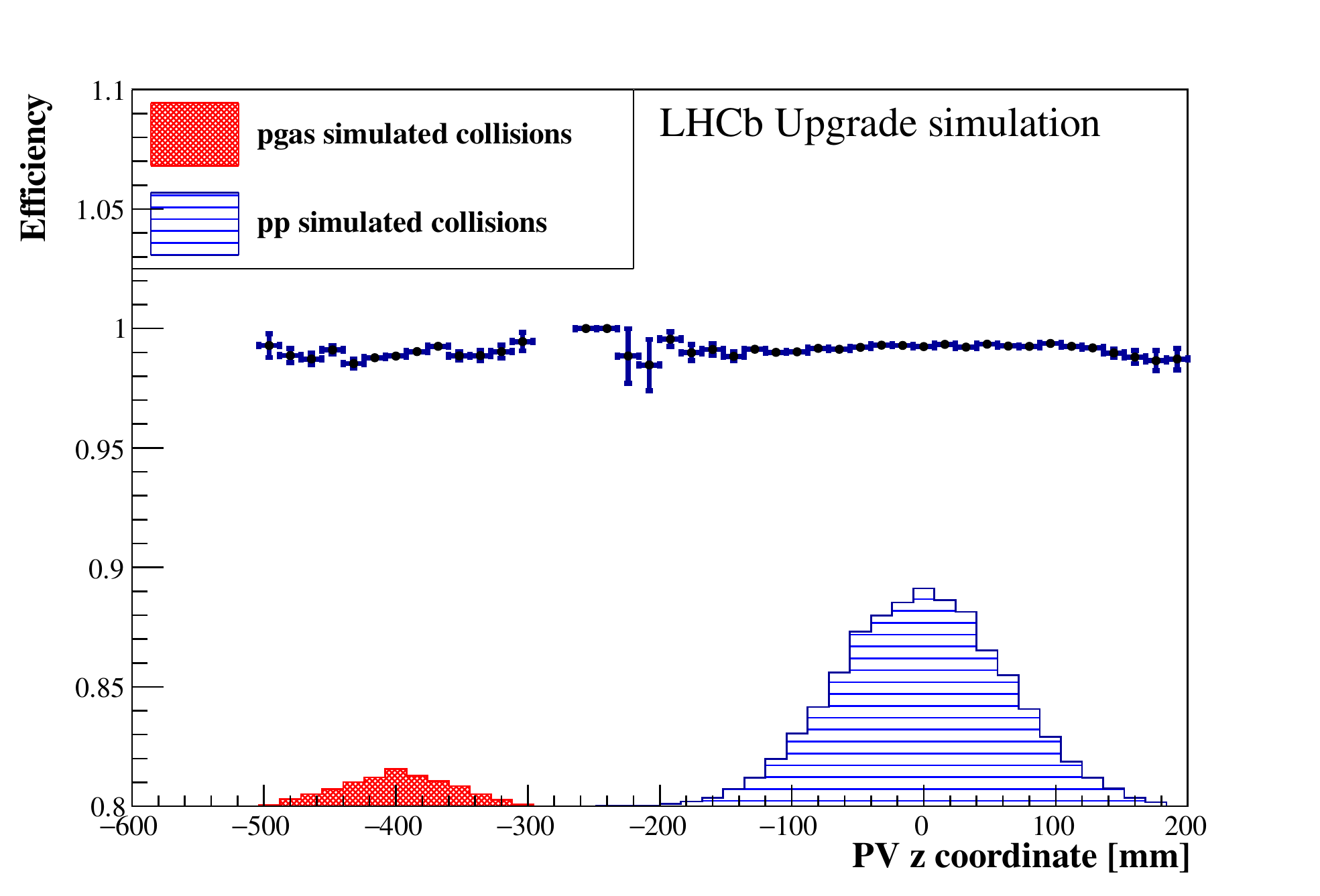}
\hfill
\includegraphics[width=0.47\textwidth]{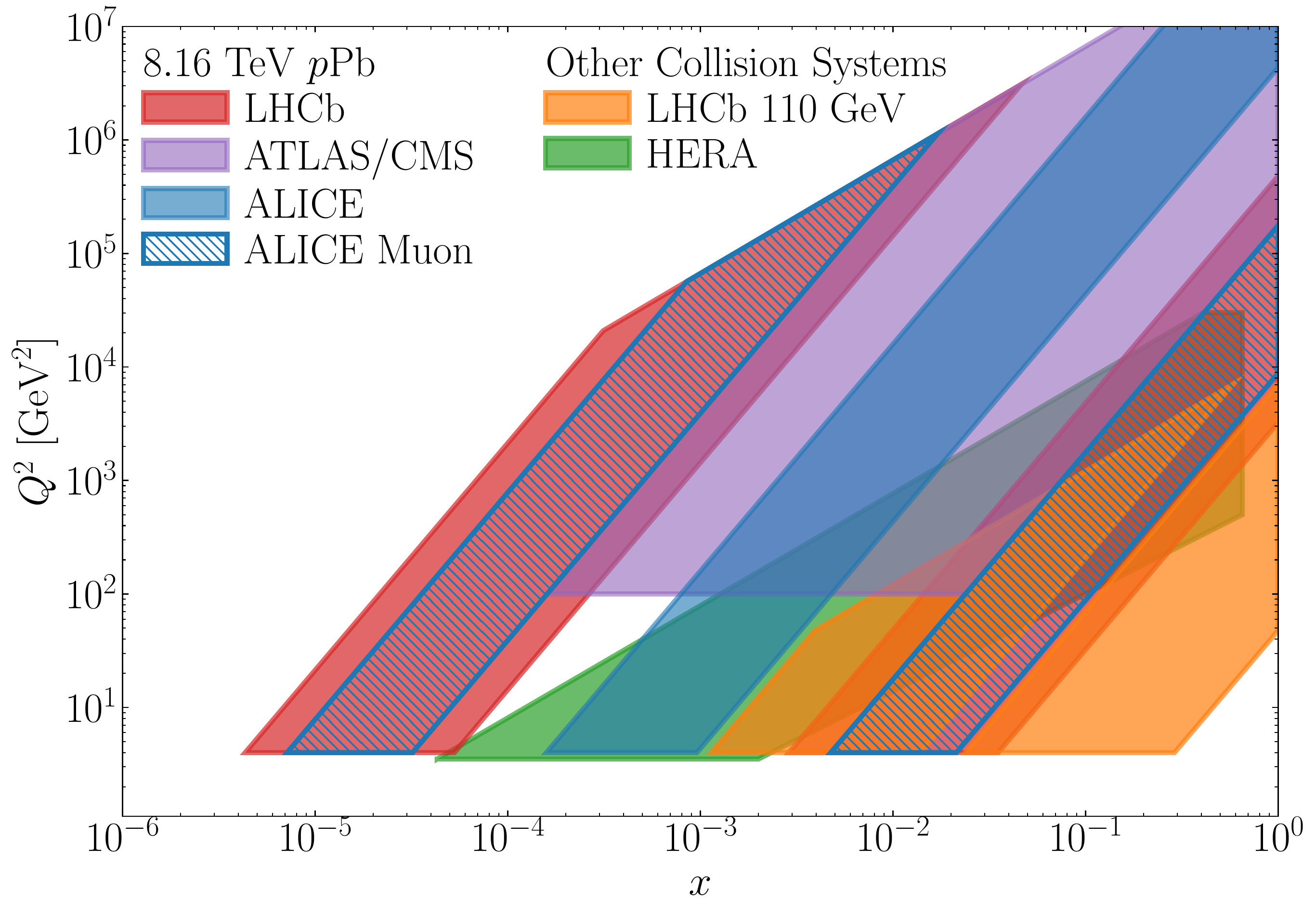}
\caption{Left: VELO track reconstruction efficiency for beam-gas (red) and beam-beam (blue) primary vertices (PV)~\cite{LHCB-FIGURE-2019-007}. Right: kinematic coverage of LHCspin (orange) and other existing facilities.}
\label{fig:kin}
\end{figure}

With the SMOG2 upgrade~\cite{LHCbCollaboration:2673690}, an openable gas storage cell, shown in Fig.~\ref{fig:smog2}, has been installed in 2020 in front of the VELO. 
The cell boosts the target areal density by a factor of $8$ to $35$ depending on the injected gas species. In addition, SMOG2 data will be collected in the upcoming Run 3 with a novel reconstruction software allowing simultaneous data-taking of beam-gas and beam-beam collisions. Very high tracking efficiency is expected in the beam-gas interaction region, despite its upstream position with respect to the VELO. Furthermore, beam-gas and beam-beam vertices are well detached along the $z$ coordinate, as shown in Fig.~\ref{fig:kin}.
SMOG2 will offer a rich physics program for the Run 3 and, at the same time, allows to investigate the dynamics of the beam-target system, setting the basis for future developments.
\\
The LHCspin project~\cite{Aidala:2019pit} aims at extending the SMOG2 physics program in Run 4 (expected to start in 2028) and, with the installation of a polarised gas target, to bring spin physics at LHC by exploiting the well suited LHCb detector. 
A selection of physics opportunities accessible at LHCspin is presented in Sec.~\ref{sec:phys}, while the experimental setup is discussed in Sec.~\ref{sec:det}.

\section{Physics case}
\label{sec:phys}
The physics case of LHCspin covers three main areas: exploration of the wide physics potential offered by unpolarised gas targets, investigation of the nucleon spin and heavy ion collisions.

\subsection{Measurements with unpolarised gases}
\label{ssec:pdfs}
\begin{figure}[ht]
\centering
\includegraphics[width=0.40\textwidth]{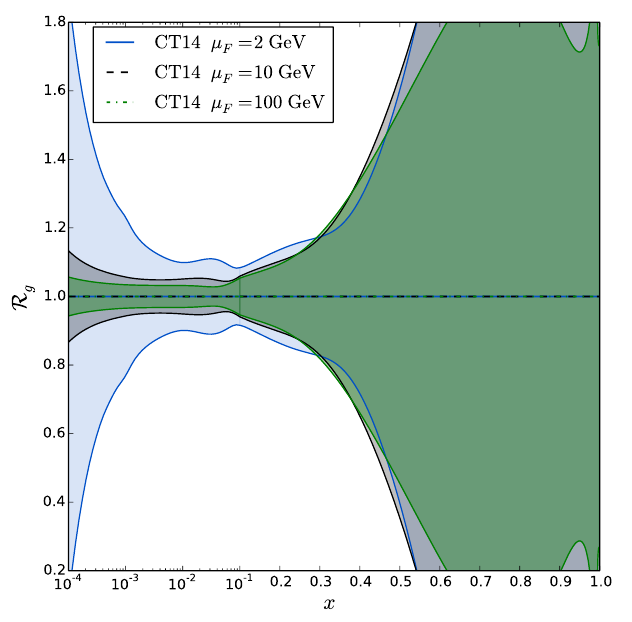}
\hfill
\includegraphics[width=0.50\textwidth]{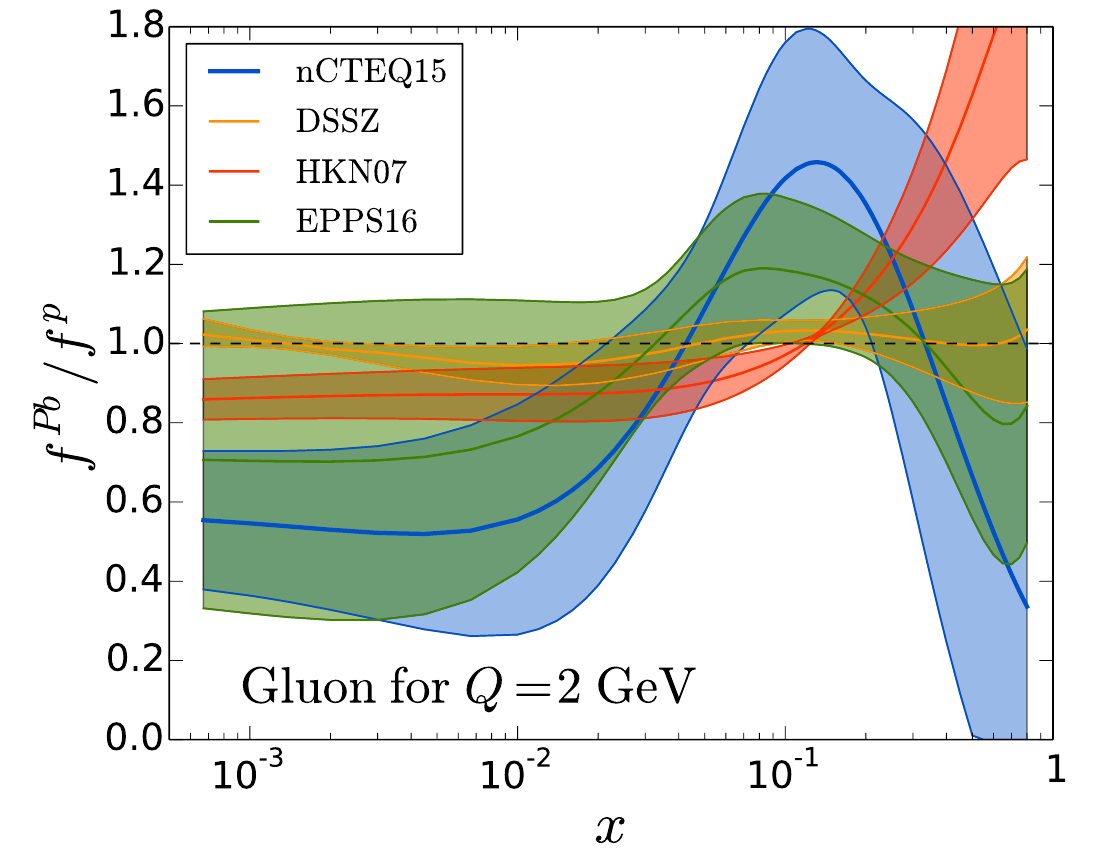}
\caption{Left: Relative uncertainty on the CT$14$ gluon PDF in a proton~\cite{Dulat:2015mca}. Right: relative uncertainties on a set of gluon nuclear PDFs in a lead nucleus~\cite{Aidala:2019pit}.}
\label{fig:gpdf}
\end{figure}

Similarly to SMOG2, LHCspin will allow the injection of several species of unpolarised gases: $\rm{H}_2$, $\rm{D}_2$, He, $\rm{N}_2$, $\rm{O}_2$, Ne, Ar, Kr and Xe. The impact of the gas on the LHC beam lifetime is negligible: the luminosity loss due to collisions on hydrogen in the cell has a characteristic time of around $2000$ days, whereas typical runs last for $10-20$ hours.
\\
Injecting unpolarised gases yields excellent opportunities to investigate parton distribution functions (PDFs) in both nucleons and nuclei in the large-$x$ and intermediate $Q^2$ regime (Fig.~\ref{fig:kin}, right), which are especially affected by lack of experimental data and impact several fields of physics from QCD to astrophysics.
For example, the large acceptance and high reconstruction efficiency of LHCb on heavy flavour states enables the study of gluon PDFs, which represent fundamental inputs for theoretical predictions~\cite{Hadjidakis:2018ifr} and are currently affected by large uncertainties, as shown in the example of Fig.~\ref{fig:gpdf} (left).
In addition, the structure of heavy nuclei is known to depart from that obtained by the simple sum of free protons and neutrons: within the unique acceptance of LHCb, a large amount of data can be collected to shed light on the intriguing anti-shadowing effect expected at $x\sim 0.1$~\cite{Eskola:2016oht}, as shown in Fig.~\ref{fig:gpdf} (right).

With the large amount of data to be collected with LHCspin, several measurements impacting astrophysics and cosmic ray physics become possible.
For example, heavy-flavour hadroproduction directly impacts the knowledge of prompt muonic neutrino flux~\cite{Garzelli:2016xmx}, which is especially affected by PDF uncertainties, while large samples of proton collisions on helium, oxygen and nitrogen provide valuable inputs to improve the understanding of the compositions of ultra-high energy cosmic rays. Moreover, the possibility of injecting an oxygen beam opens new and exciting prospects for antiproton measurements~\cite{Brewer:2021kiv}. 

\subsection{Spin physics}
\begin{figure}[ht]
\centering
\includegraphics[width=0.99\textwidth]{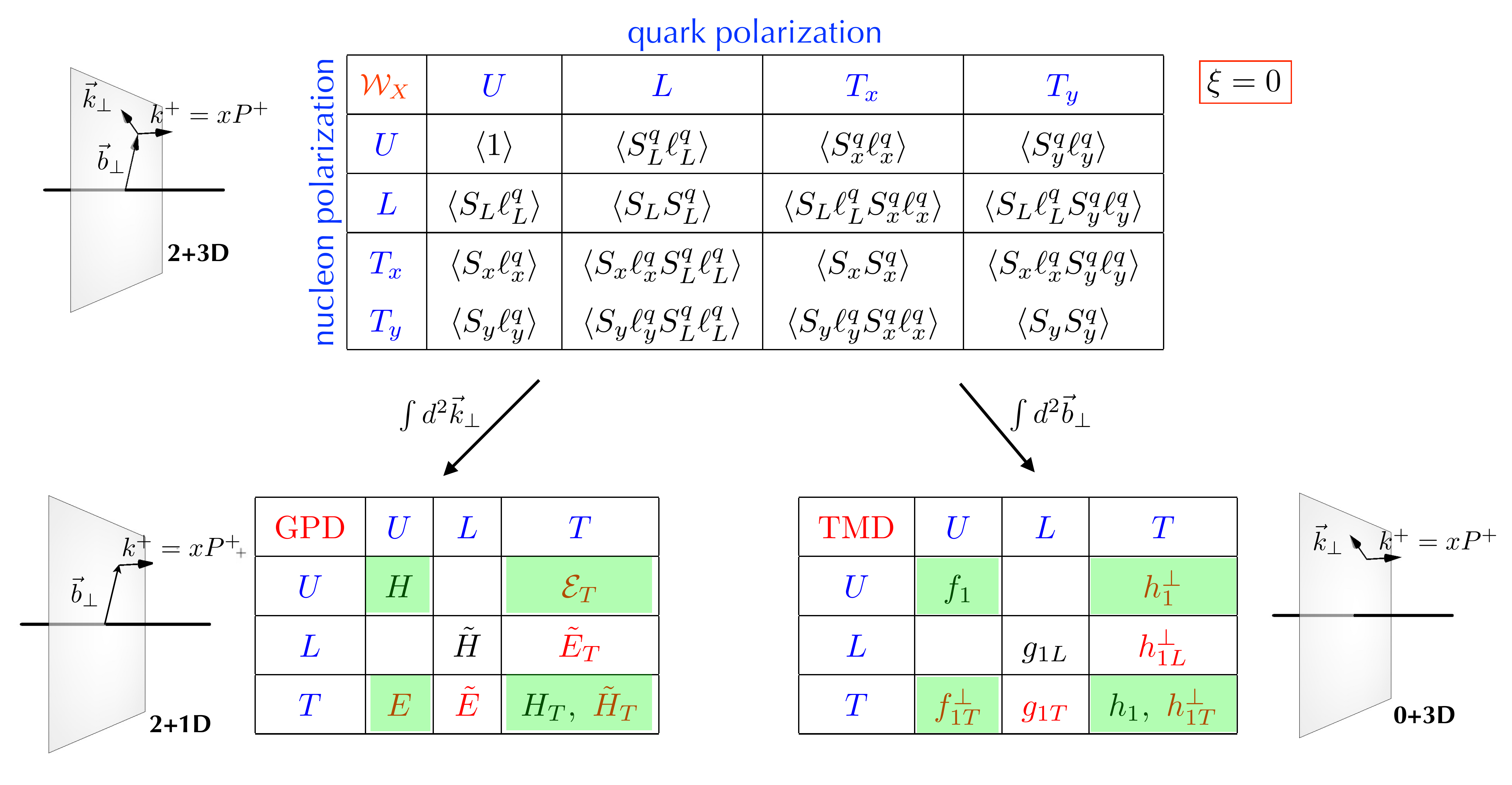}
\caption{Wigner distributions (top) and leading-twist GPDs and TMDs for different combinations of quark$\times$nucleon polarisation states (bottom). Distributions marked in red vanish for no orbital angular momentum contribution to the nucleon spin, while the quantities highlighted in green can be accessed at LHCspin~\cite{pasquini}.}
\label{fig:wigner}
\end{figure}

Beside the colliner PDFs mentioned in Sec.~\ref{ssec:pdfs}, polarised quark and gluon distributions can be probed at LHCspin by means of proton collisions on polarised hydrogen and deuterium.
Fig.~\ref{fig:wigner} shows the 5D Wigner distributions~\cite{Bhattacharya:2017bvs} which, upon integration on the transverse momentum, lead to the observable generalised parton distributions (GPDs), while transverse momentum dependent distributions (TMDs) are obtained when integrating over the transverse coordinate. There are several leading-twist distributions that can be probed with unpolarised and transversely polarised quarks and nucleons, giving independent information on the spin structure of the nucleon.

\begin{figure}[ht]
\centering
\includegraphics[width=0.56\textwidth]{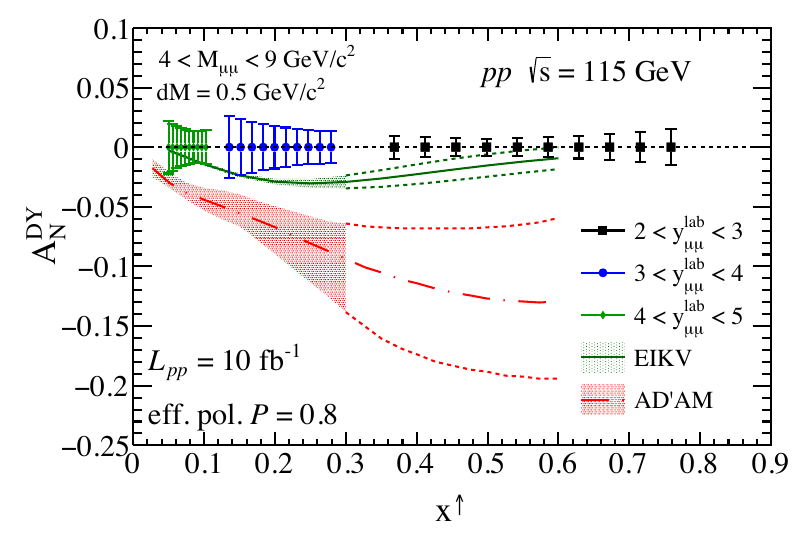}
\hfill
\includegraphics[width=0.42\textwidth]{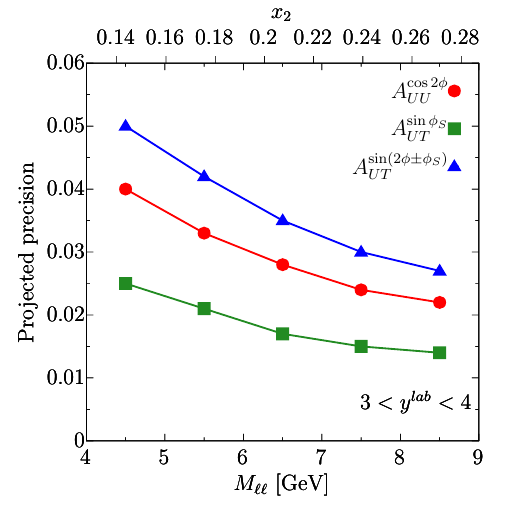}
\caption{Left: Measurements of $A_N$ in DY events as a function of $x$ compared to two theoretical predictions~\cite{Aidala:2019pit}. Right: projected precision for some azimuthal asymmetry amplitudes with DY data as a function of the dilepton invariant mass~\cite{Hadjidakis:2018ifr}.}
\label{fig:dy}
\end{figure}

To access the transverse motion of partons within a polarised nucleon, transverse single spin asymmetries (TSSAs) can be measured. For example
\begin{equation}
    A_N=\frac{1}{\it{P}}\frac{\sigma^{\uparrow}-\sigma^{\downarrow}}{\sigma^{\uparrow}+\sigma^{\downarrow}} \sim \frac{f^q_1(x_1,k^2_{T1}) \otimes f^{\perp\overline{q}}_{1T}(x_2,k^2_{T2})}{f^q_1(x_1,k^2_{T1}) \otimes f^{\overline{q}}_1(x_2,k^2_{T2})}
\end{equation}
in the polarised Drell-Yan (DY) channel probes the product of $f_1$ (unpolarised TMD) and $f^{\perp}_{1T}$ (Sivers function) in quarks and antiquarks in the low $(x_1)$ and high $(x_2)$ $x$ regimes. Projections for the uncertainty of such measurements are shown in Fig.~\ref{fig:dy} (left) based on an integrated luminosity of $10~\rm{fb}^{-1}$. Being T-odd, it is theoretically established that the Sivers function changes sign in polarised DY with respect to semi-inclusive deep inelastic scattering~\cite{Collins:2002kn}. This fundamental QCD prediction can be verified by exploiting the large sample of DY data expected at LHCspin. In addition, isospin effects can be investigated by comparing $p-\rm{H}$ and $p-\rm{D}$ collisions.
\\
Several TMDs can be probed by evaluating the azimuthal asymmetries of the produced dilepton pair: projected precisions for three such asymmetries are shown in Fig.~\ref{fig:dy} (left).

\begin{figure}[ht]
\centering
\includegraphics[width=0.49\textwidth]{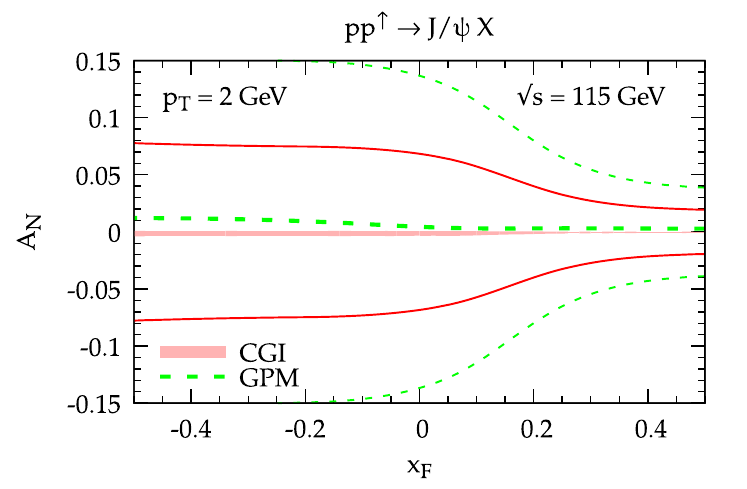}
\hfill
\includegraphics[width=0.49\textwidth]{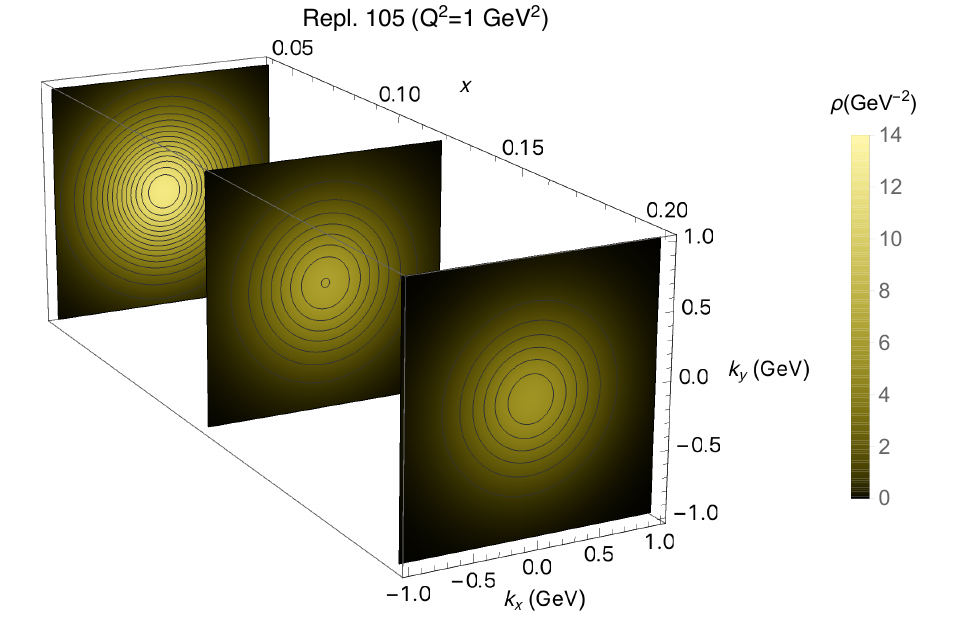}
\caption{Left: theoretical predictions for $A_N$ in inclusive $J/\psi$ production~\cite{DAlesio:2018rnv}. Right: up quark densities in momentum space~\cite{Bacchetta:2017gcc}.}
\label{fig:tmds}
\end{figure}

Heavy flavour states will be the strength point of LHCspin. Being mainly produced via gluon fusion at LHC, quarkonia and open heavy flavour states will allow to probe the unknown gluon Sivers function via inclusive production of $J/\psi$, $D^0$ but also with several unique states like $\eta_c$, $\chi_c$, $\chi_b$ or $J/\psi J/\psi$.
Fig.~\ref{fig:tmds} (left) shows two predictions for $A_N$ on $J/\psi$ events: $5-10\%$ asymmetries are expected in the $x_F<0$ region, where the LHCspin sensitivity is the highest.
Heavy flavour states can be exploited as well to probe the gluon-induced asymmetries $h^{\perp g}_{1}$ (Boer-Mulders) and $f_1^g$ (always present at the denominator of $A_N$), which are both experimentally unconstrained. 

\begin{figure}[ht]
\centering
\includegraphics[width=0.8\textwidth]{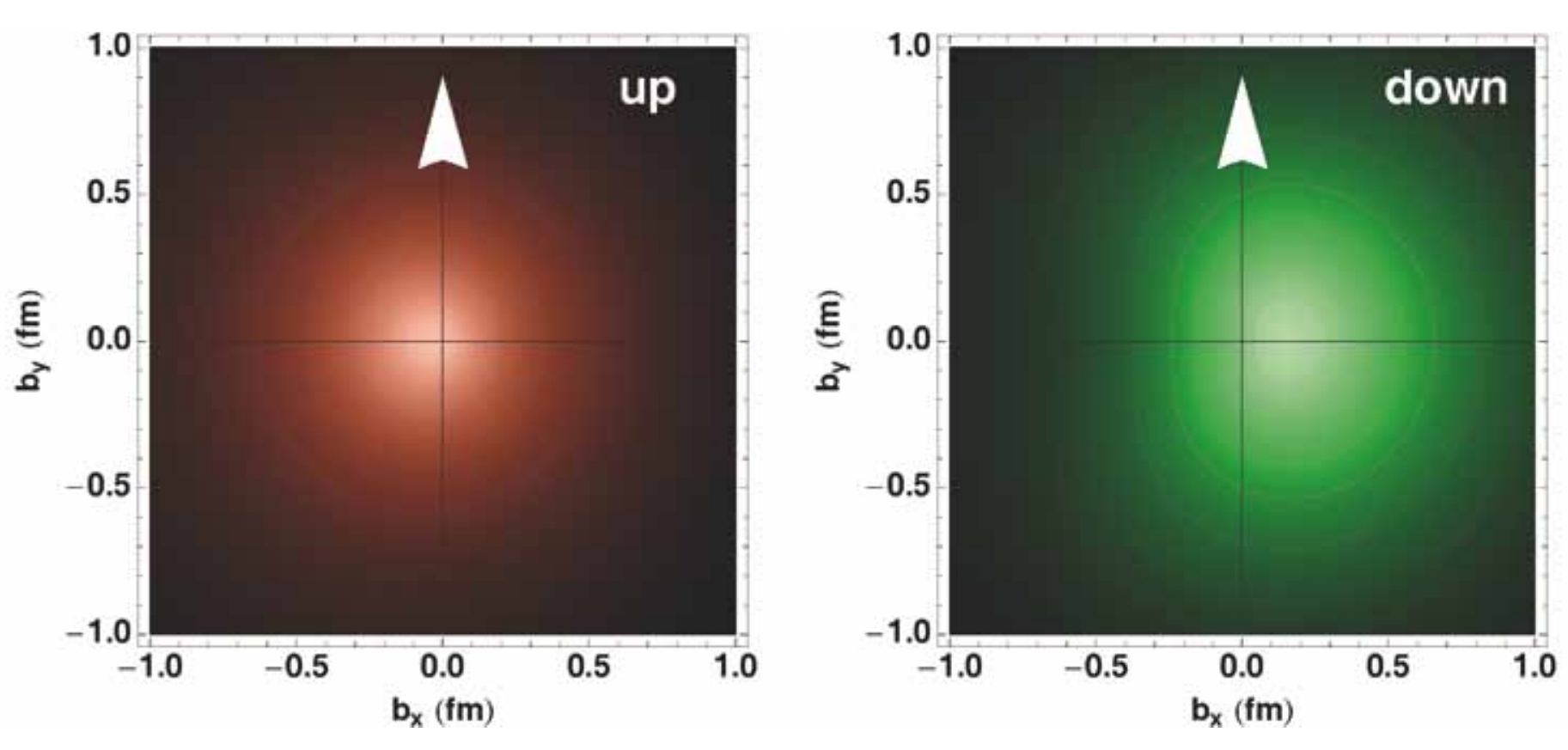}
\caption{Distortion of the up and down quark distributions in the impact parameter space when spin is taken into account~\cite{saggiatore}.}
\label{fig:pic}
\end{figure}

While TMDs provide a ``tomography'' of the nucleon in momentum space (Fig.~\ref{fig:tmds}, right), a 3D picture in the spatial coordinates can be built by measuring GPDs, as shown in Fig.~\ref{fig:pic}.
Correlating position and momentum, GPDs also quantify the parton orbital angular momentum, whose contribution to the total nucleon spin can be inferred, for example, via the Ji sum rule~\cite{Ji:1996ek}.
GPDs can be experimentally probed with exclusive dilepton and quarkonia productions in ultra-peripheral collisions, which are dominated by the electromagnetic interaction.
For example, TSSAs can be exploited to access the $E_g$ function, which has never been measured and represents a key element of the proton spin puzzle.
It is also attractive to measure the elusive
transversity PDF, whose knowledge is currently limited to valence quarks at the leading order~\cite{Radici:2018iag}, as well as its integral, the tensor charge, which is of direct interest in constraining physics beyond the Standard Model~\cite{Courtoy:2015haa}. 

\subsection{Heavy ion collisions}
\begin{figure}[h]
\centering
\includegraphics[width=0.47\textwidth]{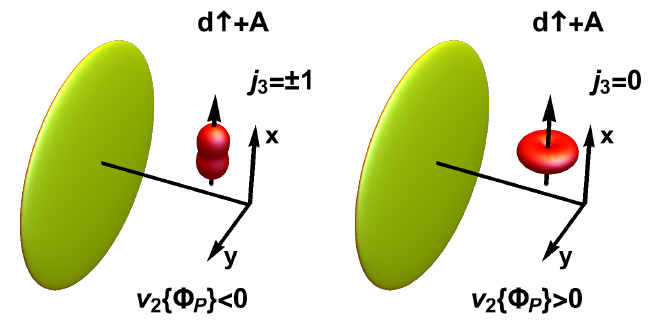}
\hspace{1cm}
\includegraphics[width=0.42\textwidth]{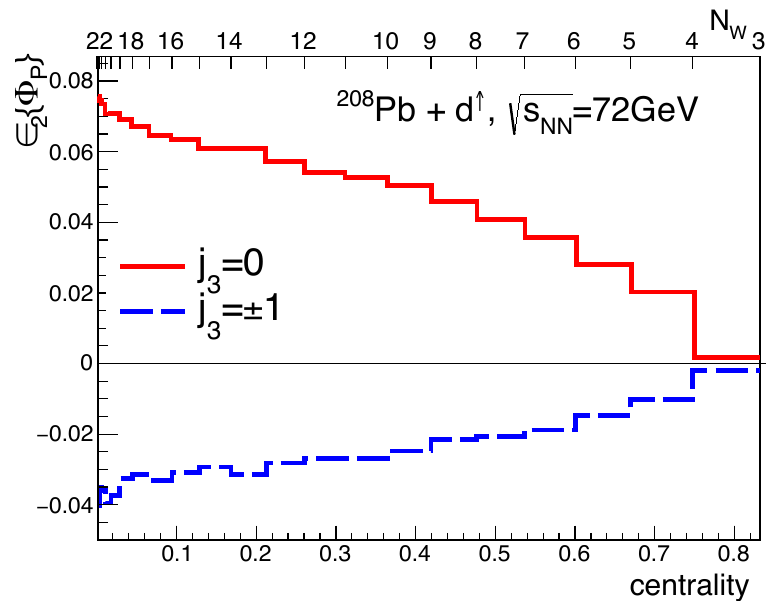}
\caption{Left: sketch of a ultra-relativistic collision of a lead nucleus against a transversely polarised deuteron in two different angular momentum projections. Right: ellipticity with respect to the polarisation axis as a function of the collision centrality with LHCspin kinematics~\cite{Broniowski:2019kjo}.}
\label{fig:deuteron}
\end{figure}

Thermal heavy-flavour production is negligible at the typical temperature of few hundreds MeV of the system created in heavy-ion collisions. Quarkonia states ($c\overline{c}$, $b\overline{b}$) are instead produced on shorter timescales, and their energy change while traversing the medium represents a powerful way to investigate Quark-Gluon Plasma (QGP) properties. LHCb capabilities allow to both cover the aforementioned charmonia and bottomonia studies and to extend them to beauty baryons as well as to exotic probes.
The QGP phase diagram exploration at LHCspin can be performed with a rapidity scan at a centre of mass energy of $\sqrt{s_{\rm{NN}}}=72~\rm{GeV}$ which is in-between those accessed at RHIC and SPS. 
In addition, flow measurements will greatly benefit from the excellent identification performance of LHCb on charged and neutral light hadrons.
\\
The dynamics of small systems is an interesting topic joining heavy-ion collisions and spin physics.
In the spin $1$ deuteron nucleus, the nucleon matter distribution is prolate for $j_3=\pm1$ and oblate for $j_3=0$, where $j_3$ is the projection of the spin along the polarisation axis.
In ultra-relativistic lead ion collisions on transversely polarised deuteron, the deformation of the target deuteron can influence the orientation of the fireball in the transverse plane, quantified by the ellipticity, as shown in Fig.~\ref{fig:deuteron}.
The measurement proposed in~\cite{Broniowski:2019kjo} can easily be performed at LHCspin on minimum bias events thanks to the high-intensity LHC beam.

\section{Experimental setup}
\label{sec:det}
\begin{figure}[ht]
\centering
\includegraphics[width=0.9\textwidth]{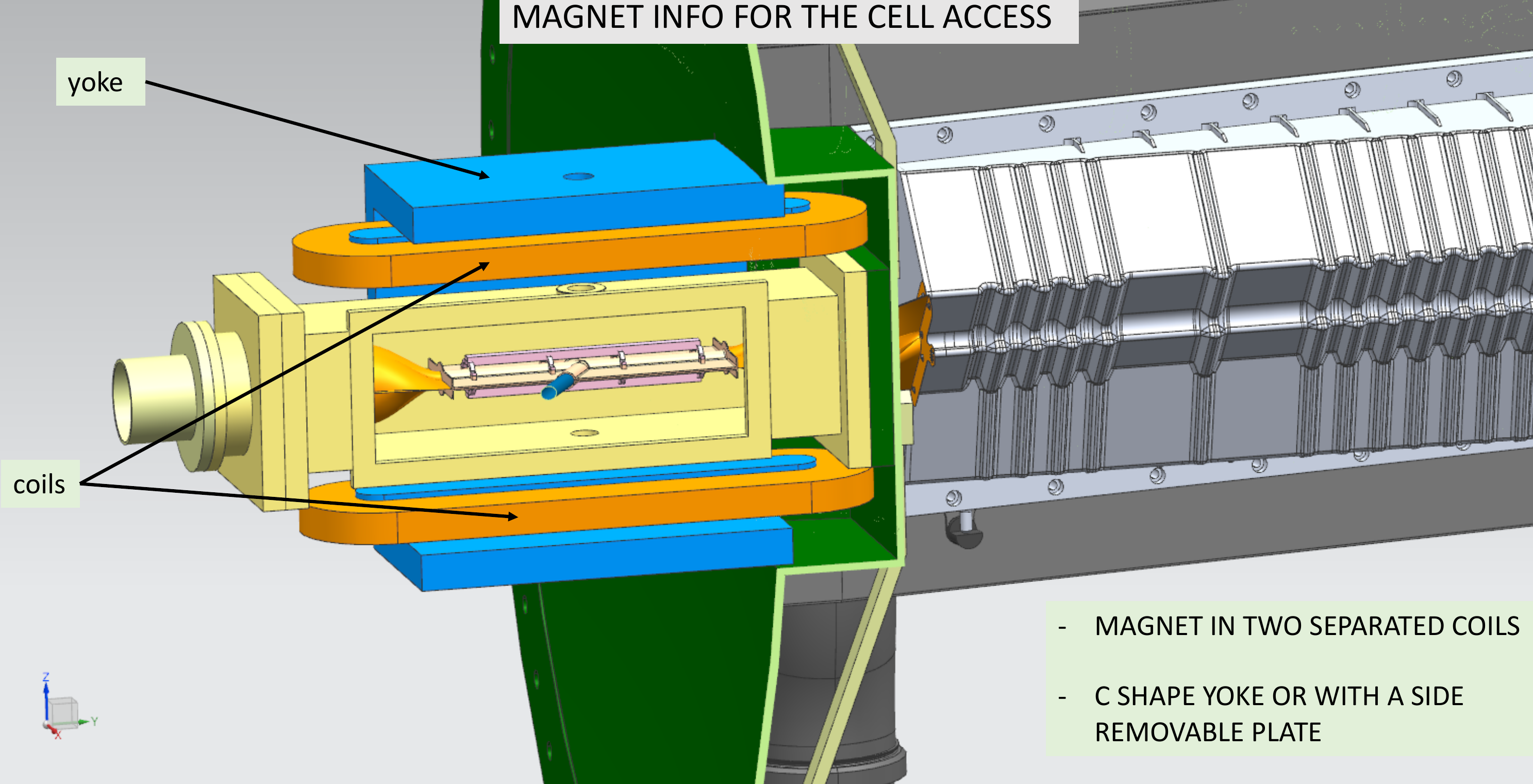}
\caption{A drawing of the PGT with the magnet coils (orange) and the iron return yoke (blue) enclosing the storage cell. The VELO vessel and detector box are shown in green and grey, respectively.}
\label{fig:rd}
\end{figure}

The LHCspin experimental setup is in R\&D phase and calls for the development of a new generation polarised target. The starting point for this ambitious task is the setup of the polarised target system employed at the HERMES experiment~\cite{Airapetian:2004yf} and comprises three main components: an Atomic Beam Source (ABS), a Polarised Gas Target (PGT) and a diagnostic system.
The ABS consists of a dissociator with a cooled nozzle, a Stern-Gerlach apparatus to focus the wanted hyperfine states, and adiabatic RF-transitions for setting and switching the target polarisation between states of opposite sign.
The ABS injects a beam of polarised hydrogen or deuterium into the PGT, which is located in the LHC primary vacuum.
The PGT hosts a T-shaped openable storage cell, sharing the SMOG2 geometry, and a compact superconductive dipole magnet, as shown in Fig.~\ref{fig:rd}. The magnet generates a $300~\rm{mT}$ static transverse field with a homogeneity of 10\%, which is found to be suitable to avoid beam-induced depolarisation~\cite{Steffens:2019kgb}.
Studies for the inner coating of the storage cell are currently ongoing, with the aim of producing a surface that minimises the molecular recombination rate as well as the secondary electron yield.

\begin{figure}[ht]
\centering
\includegraphics[width=0.49\textwidth]{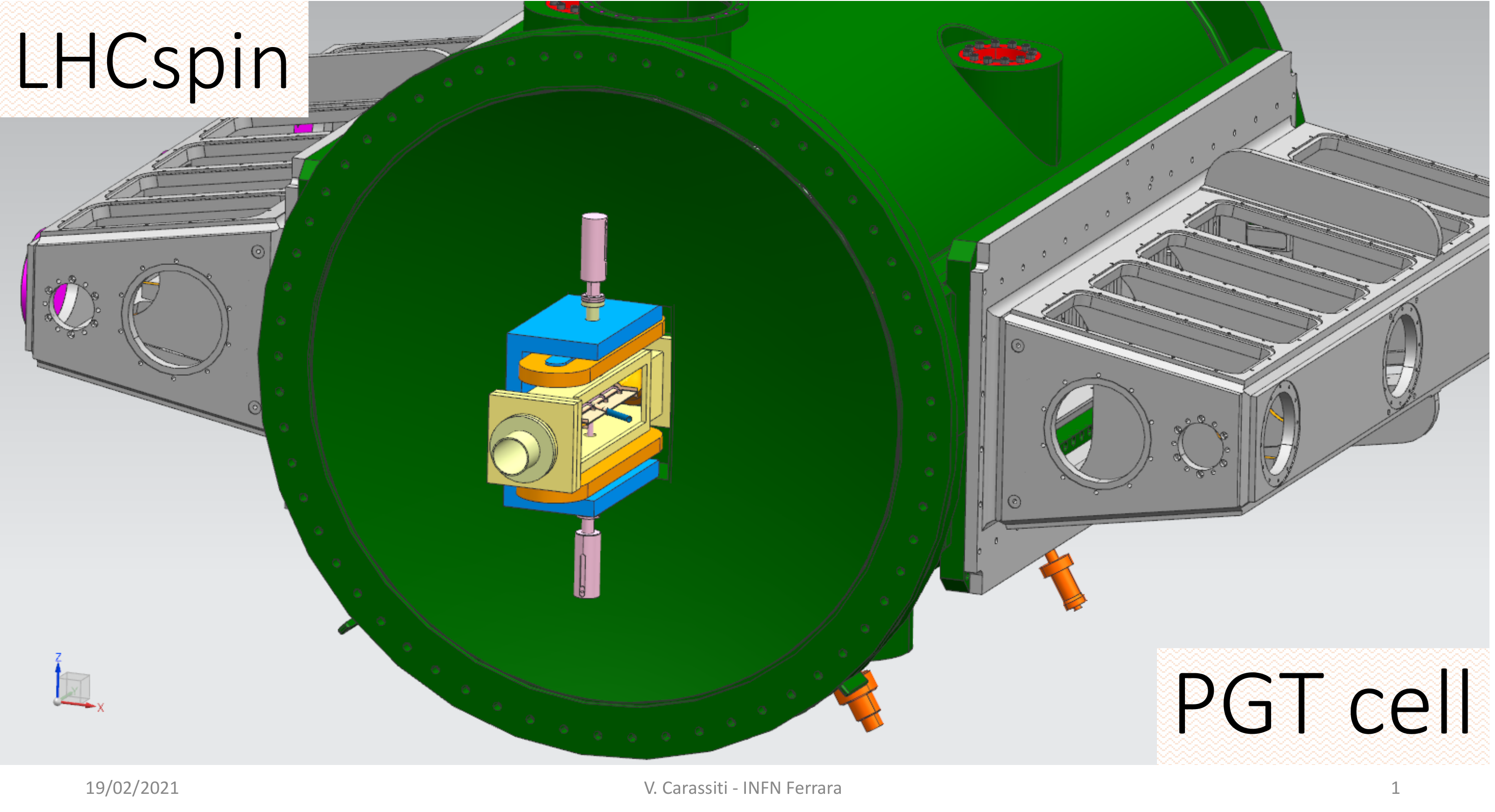}
\hfill
\includegraphics[width=0.45\textwidth]{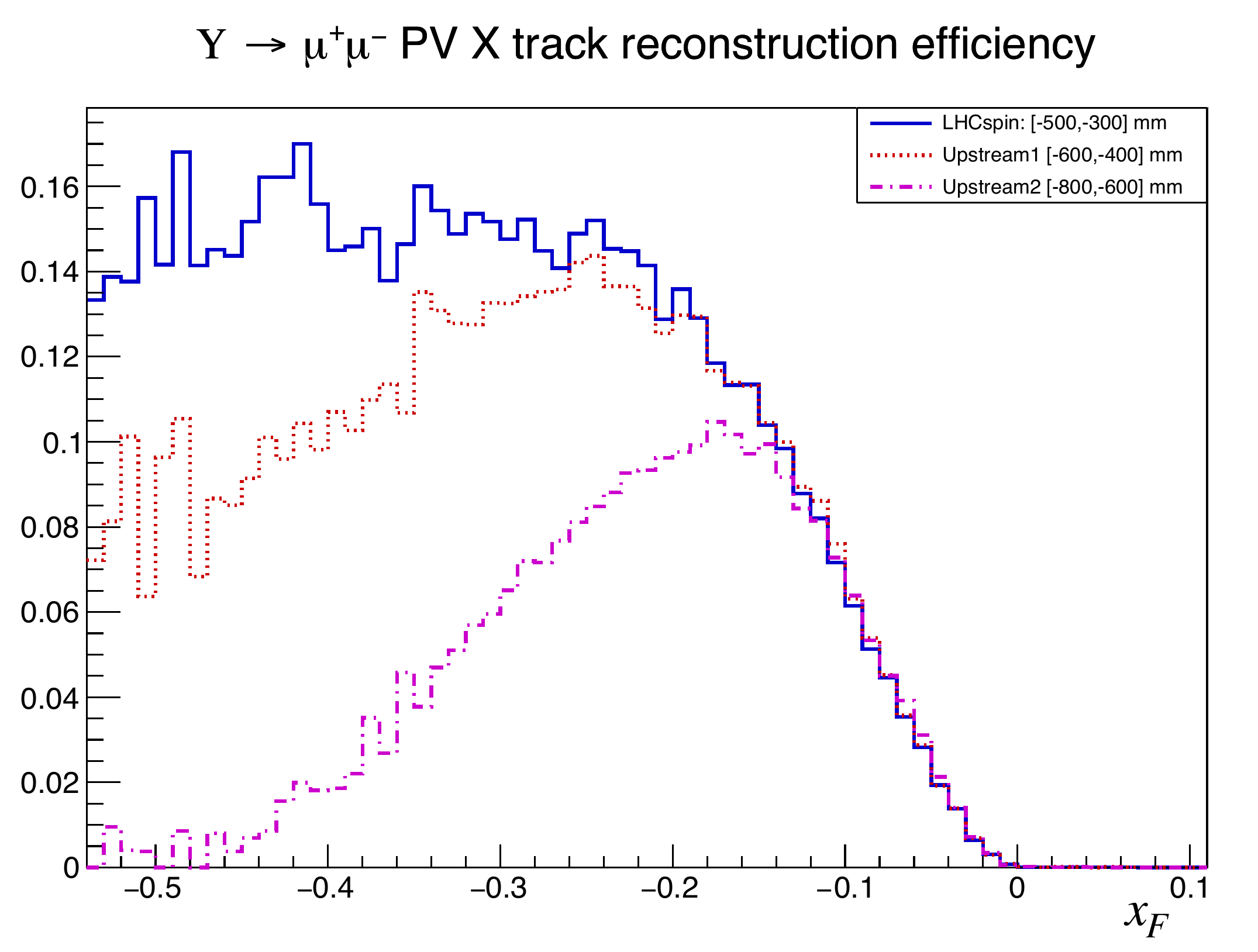}
\caption{Left: The PGT and the VELO vessel (green). Right: simulated reconstruction efficiency for $\Upsilon\to\mu^+\mu^-$ events with three different cell positions, the blue line corresponding to the SMOG2 location.}
\label{fig:pgtvelo}
\end{figure}

In Fig.~\ref{fig:pgtvelo} (left), the PGT is shown in the same location of the SMOG2 cell, a configuration that offers a large kinematic acceptance and does not require additional detectors in LHCb. Fig.~\ref{fig:pgtvelo} (right) shows the efficiency to reconstruct a primary vertex and both tracks in simulated \mbox{$\Upsilon\to\mu^+\mu^-$} events as a function of $x_F$ under three locations of a $20~\rm{cm}$-long storage cell.
The simulation is performed within the GAUSS framework~\cite{Clemencic:2011zza} with upgrade LHCb conditions. New algorithms are currently being developed for the Run 3 fixed-target reconstruction and are expected to sensibly improve the current performance as well as to enable to record LHCspin data in parallel with $p-p$ collisions~\cite{LHCB-FIGURE-2019-007}.

The diagnostic system continuously analyses gas samples drawn from the PGT and comprises a target gas analyser to detect the molecular fraction, and thus the degree of dissociation, and a Breit-Rabi polarimeter to measure the relative population of the injected hyperfine states.
\\
An instantaneous luminosity of $\mathcal{O}(10^{32})~\rm{cm}^{-2}\rm{s}^{-1}$ is foreseen for fixed-target $p-\rm{H}$ collisions in Run 4, with a further factor $3-5$ increase for the high-luminosity LHC phase from Run 5 (2032). 

\section{Conclusions}
The fixed-target physics program at LHC has been greatly enhanced with the recent installation of the SMOG2 gas storage cell at LHCb.
LHCspin is the natural evolution of SMOG2 and aims at installing a polarised gas target to bring spin physics at LHC for the first time, opening a whole new range of exploration. With strong interest and support from the international theoretical community, LHCspin is a unique opportunity to advance our knowledge on several unexplored QCD areas, complementing both existing facilities and the future Electron-Ion Collider~\cite{Accardi:2012qut}.

\paragraph{Funding information}
This project has received funding from the European Union’s Horizon 2020 research and innovation programme under grant agreement STRONG – 2020 - No 824093.

\bibliographystyle{JHEP} 
\bibliography{main.bib}

\end{document}